\documentstyle[graphicx,floats,prl,aps]{revtex}

\tighten
\draft

\begin{document}

\twocolumn[\hsize\textwidth\columnwidth\hsize\csname
@twocolumnfalse\endcsname
\title{How magic is the magic  $^{68}$Ni nucleus?}
\author {K. Langanke$^{1,2}$, 
 J. Terasaki$^{2-4}$, F. Nowacki$^5$,
 D.J. Dean$^2$, and W. Nazarewicz$^{2,3,6}$}
\address{
$^1$Institute of Physics and Astronomy, University of Aarhus,
DK-8000 Aarhus C, Denmark \\
$^2$Physics Division, 
Oak Ridge National Laboratory, Oak Ridge, TN
37831 USA \\
$^3$Department of Physics, University of Tennessee, Knoxville, TN
37996, USA \\
$^4$Joint Institute for Heavy Ion Research, Oak Ridge National Laboratory, Oak Ridge, TN
37831 USA \\
$^5$Institut de Recherches Subatomiques, IN2P3-CNRS-Universit\'e Louis
Pasteur, F--67037 Strasbourg Cedex~2, France \\
$^6$Institute for Theoretical Physics, University of Warsaw,
ul. Ho\.za 69, PL-00-681, Warsaw, Poland
}
\date{\today}
\maketitle

\begin{abstract}
We calculate the $B(E2)$
strength in $^{68}$Ni and other nickel isotopes using several theoretical
 approaches. We find that in
$^{68}$Ni the gamma transition to the first $2^+$ state exhausts only a
fraction of the total $B(E2)$ strength, which is mainly collected
in excited states around 5 MeV. This effect is sensitive
to the energy splitting between the $fp$ shell and the $g_{9/2}$
orbital. We argue that the small experimental $B(E2)$ value is not  strong
evidence for the double-magic character of $^{68}$Ni.
\end{abstract}


\vspace{0.4cm}]
\narrowtext

The appearance of shell gaps associated with magic nucleon numbers is
one of the cornerstones of nuclear structure. The presence of magic
gaps allows one, for example,
to determine the single-particle energies and the residual interaction
among valence nucleons, providing  essential input for nuclear
models. Magic gaps offer a natural way of performing 
truncations in  microscopic many-body calculations. 
Magic nuclei also play  an essential role in the two major
nucleosynthesis networks (s- and r-process) that  produce the majority
of nuclides heavier than mass number $A \sim 60$. 

The doubly-magic character of $^{68}$Ni ($Z$=28, $N$=40) was suggested in 
the early eighties \cite{bernas,lombard}
and tested experimentally 
\cite{Grzywacz98,Mueller00,Ishii02,Sorlin02}. The
proton number $Z$=28 in the nickel isotopes is magic.
In the neutrons, the sizeable energy gap at $N$=40 
separates the $pf$ spherical shell from the $g_{9/2}$ intruder orbit.
However, this spherical subshell closure
is not sufficiently large to stabilize the spherical shape.
[Experimentally \cite{lister},  $^{80}$Zr ($N$=$Z$=40)
behaves like  a well-deformed rotor.]
The current experimental evidence about the double-magicity of $^{68}$Ni
is controversial \cite{Grawe01}. On the one hand,
$^{68}$Ni does not show a pronounced irregularity  in the
two-neutron separation energies, as is expected for a magic nucleus. 
On the other hand, the lowered position of the $0^+_2$ level, 
the slightly elevated energy of the first $2^+$ state, and the quite small
$B(E2, 0^+_{\rm g.s.} \rightarrow 2^+_1)$ value are often interpreted
as indications for magicity. In fact, the $B(E2)$ rate in
$^{68}$Ni ($280\pm60$ e$^2$fm$^4$ \cite{Sorlin02}) is significantly
smaller than that in the well-established double-magic nucleus $^{56}$Ni
($620\pm120$ e$^2$fm$^4$ \cite{ni56}).

As discussed in Ref.~\cite{reinhard}, the size of the $N$=40 gap strongly
depends on the effective interaction used, and it dramatically influences
the quadrupole collectivity of the $N$=40 nuclei. While there is much discussion
in the literature about the weakening of shell effects in neutron-rich nuclei
(e.g., the magic gap $N$=28 seems to be  eroded 
 in drip-line systems; see Ref.~\cite{reinhard}
 and references quoted therein), $^{68}$Ni lies  very far from the expected
 neutron drip line (expected to be around $^{92}$Ni \cite{nazarewicz}) and one should
 probably not invoke ``exotic" explanations 
  when discussing the stucture of
 this  neutron rich nucleus.

It is the aim of this Letter to draw attention to the total
$B(E2;0^+_{\rm g.s.}\rightarrow 2^+_f)$ distribution in $^{68}$Ni, 
which can hold the key to the
question whether this nucleus is magic or not. We will argue that
the 
transition to the first excited $2^+$ state constitutes only a small part of
the total $B(E2)$ strength and that the total $B(E2)$ strength
depends sensitively
on the size of the $N=40$ shell gap. 

To understand
the structural difference between $^{68}$Ni and 
 $^{56}$Ni (where the transition to the first $2^+$
state exhausts most of the total low-energy strength),  
we begin from  qualitative arguments based on
a simple independent particle model (IPM).
Proton configurations in both nuclei are
identical and correspond to a closed $(f_{7/2})$ shell. 
The neutron configuration in $^{56}$Ni is the same as the proton one, 
$(f_{7/2})^8$,  while in $^{68}$Ni it corresponds to the closed
$(pf)$ shell.
The first excited $2^+$ state 
can  be viewed  as a superposition of 
particle-hole (ph) excitations. In $^{56}$Ni, the lowest
 proton and neutron ph excitations are identical
 (due to isospin symmetry) and they are of (1p-1h) character. 
 The $2^+_1$ state can be roughly
 expressed as $1/\sqrt{2} (\Phi_p (1p-1h) + \Phi_n (1p-1h))$. 
 Due to the parity change between  $(pf)$ and  $g_{9/2}$ orbits, 
 the $2^+_1$ state in $^{68}$Ni cannot have a 1p-1h
neutron component. It can thus  be written as
$\alpha \Phi_p (1p-1h) + \beta \Phi_n (2p-2h)$, where
$|\alpha|^2+|\beta|^2$=1.
The weight $|\beta|^2$ of the neutron component decreases 
with the size of the $N$=40 gap.
As  $B(E2)$
transitions reflect a proton component in the wave function, 
an increasing value of $\beta$ reduces
the $B(E2)$ rate. Consequently, a  smaller
 $B(E2, 0^+_{\rm g.s.} \rightarrow
2^+_1$) value in $^{68}$Ni than in $^{56}$Ni  suggests that
$|\alpha|$ is noticeably smaller than $1/\sqrt{2}$, implying that it   
is more favorable in $^{68}$Ni to excite the pair of
 neutrons into the $g_{9/2}$ orbital than to
excite a single proton across the magic $N=28$ gap.
As the proton configurations in the 
$^{56}$Ni and $^{68}$Ni ground states
are the same,
the  IPM  also suggests that a noticeable part of the
(1p-1h) proton excitations in $^{68}$Ni
 should be mixed into excited states.
 Of course, the neutron amplitude $|\beta|$, hence the $B(E2)$ rate, should also
 depend on the residual interaction. Therefore, to check whether the
 simple IPM argument also holds in the presence of a realistic residual interaction,
 we performed calculations in three different theoretical models:
the Shell Model Monte Carlo (SMMC), the 
Quasi-particle Random-Phase Approximmation
(QRPA), and  a large-scale diagonalization
shell model (SM).

The SMMC approach allows the calculation of nuclear properties as
thermal averages, employing the Hubbard-Stratonovich transformation to
rewrite the two-body parts of the residual interaction by integrals over
fluctuating auxiliary fields \cite{Koonin97}. We  performed SMMC
studies of the even-even nickel isotopes between $^{56}$Ni and $^{78}$Ni
in the complete $(fp)(gds)$ configuration space  for both protons and neutrons. 
The single-particle energies were
derived from a Woods-Saxon potential appropriate for $^{56}$Ni, placing
the important levels at excitation energies (in MeV) of 
4.3 ($p_{3/2})$, 6.4 ($f_{5/2}$), 6.6
($p_{1/2})$, 9.0 ($g_{9/2})$, 13.0 ($d_{5/2}$) relative to the $f_{7/2}$
orbital.  
We employed the same residual interaction of the type
pairing-plus-quadrupole as in Ref.~\cite{Dean02}, which allowed us to  
avoid the  sign problem 
in SMMC calculations \cite{Lang93}. 
We checked that our interaction gives a reasonable description of
the collective spectrum of $^{64}$Ge and $^{64}$Ni,
and that center-of-mass contaminations are small and do not
affect our results for quadrupole excitations.
The SMMC calculations were performed at temperature 
$T$=0.33 MeV (corresponding to 96 $\Delta \beta$ `time slices'), which,
for even-even nuclei, is sufficiently low to cool the nucleus to the
ground state. We  checked this for $^{68}$Ni and found variations of
the various quadrupole expectation values of less than $3\%$ by slightly
increasing the temperature to $T$=0.4 MeV. 
The Monte Carlo integrations used between 1000 and 4000 samples.

In the SMMC approach, the total $B(E2)$ strength
is  approximated by   the 
expectation value  $B(E2) \propto \langle \hat{Q}^2 \rangle$, where
the quadrupole operator is defined by $\hat{Q}= e_p \hat{Q}_p + 
e_n \hat{Q}_n$, with
$\hat{Q}_{p(n)} = \sum_i r_i^2 Y_2 (\theta_i,\phi_i)$; the sum runs over all
valence protons (neutrons). The effective charges $e_p,e_n$ account for
coupling to the states outside  our model space.
We adopt in the following the standard values $e_p=1.5, e_n=0.5$
\cite{Martinez94}. For the single-particle wave functions, we adopt
the harmonic oscillator states with the oscillator length 
$b=1.01 A^{1/6}$ fm.
At low temperatures, the total $B(E2)$ strength obtained in SMMC approximates
the expectation value $\langle 0^+_{\rm g.s.}|\hat{Q}^2 |0^+_{\rm g.s.}\rangle=
\sum_f |\langle 0^+_{\rm g.s.}|\hat{Q}|2^+_f\rangle|^2 \propto  \sum_f
B(E2; 0^+_{\rm g.s.} \rightarrow 2^+_f)$. Therefore, the total SMMC strength
corresponds to the {\it summed} $B(E2)$ strength to the excited $2^+$ states 
within the assumed configuration space.

\begin{figure}[ht]
  \begin{center}
    \leavevmode
    \includegraphics[width=0.80\columnwidth]{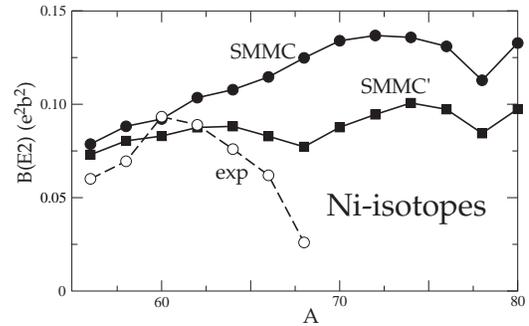}
    \caption{
Comparison of the total SMMC $B(E2)$ values for even-even nickel 
isotopes (solid circles) with the experimental $B(E2, 0^+_{\rm g.s.} \rightarrow
2^+_1)$  rates (open circles, from Ref.~\protect\cite{Raman}). The solid squares
represent the SMMC $B(E2)$ values obtained 
in the SMMC' variant of calculations in which  all the $gds$ single-particle
energies are shifted up by 1 MeV.
        }
    \label{fig:smmc}
  \end{center}
\end{figure}

The total SMMC $B(E2)$ strengths are plotted in Fig.~\ref{fig:smmc}.
They follow the experimental transition  rates rather
closely up to $^{62}$Ni. For these nuclei, it is well known from electron
scattering experiments that most of the $B(E2)$ strength resides  in the
transition to the first $2^+$ state (see, e.g., Ref.~\cite{Langanke95}).
For the isotopes approaching  $N=40$,  SMMC predicts a significantly
larger total $B(E2)$ strength than observed in the first transition. 
For $^{68}$Ni,  the calculated  total strength ($\sim 1250 e^2{\rm fm}^4$) is
about five times greater than
the measured transition to the first $2^+$ state. This is
consistent with the fact that 
the centroid of the SMMC $B(E2)$ strength, calculated from the
respective response function, lies at $\sim 5$ MeV. This value is significantly
higher than the energy of the  first $2^+$ state and  indicates that 
most of the calculated  SMMC $B(E2)$ strength in $^{68}$Ni
resides  in excited states. 

In Ref.~\cite{Sorlin02} it was pointed out that the neutron $g_{9/2}$
orbital plays a major role at $N=40$,  thanks to
cross-shell neutron pairing excitations. 
We confirm this finding. In our SMMC study we find an average occupation
number $\langle n \rangle$=2.2 for  neutrons 
in the $g_{9/2}$ orbital (and 0.22 in the
$d_{5/2}$ orbital which couples strongly to $g_{9/2}$ by the quadrupole
force). These numbers are significantly reduced
(to 0.9 and 0.08, respectively)
 if one  artificially shifts upwards
all levels of the $gds$ shell by 1 MeV, making the $N$=40 gap
larger (see the SMMC' variant of calculations in Fig.~\ref{fig:smmc}). 
 As a consequence, the
quadrupole moment  of the neutron configuration gets reduced and 
 the summed  $B(E2)$ value decreases. 
 Although in this modified calculation
the summed $B(E2)$ value  is smaller for $^{68}$Ni than in the neighboring
nuclei, our calculation still predicts  most of the strength in
excited states.
We note that for both sets of single-particle energies,  $^{78}$Ni is
characterized as magic, having a reduced $B(E2)$ value compared to the
neighbors. It is also worth mentioning that the variations of $Q_p^2$
along the isotope chain are rather small. The smallest value is obtained  for
$^{56}$Ni, the largest for $^{72}$Ni; however, the variation is less
than $9\%$. This shows again that
proton ($f_{7/2}   \rightarrow 
p_{3/2},p_{1/2},f_{5/2}$) excitations cannot be neglected
in the reproduction of the $E2$
 strength, as already noted in \cite{Sorlin02}, 
but the dominating variations in the $B(E2)$ strength  come from the 
neutrons.

\begin{figure}[ht]
\begin{center}
    \leavevmode
    \includegraphics[width=0.80\columnwidth]{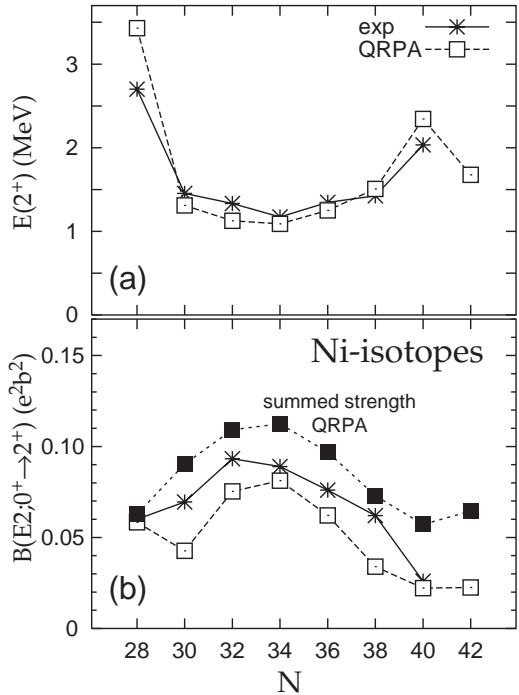}
\caption{(a) Comparison of the QRPA $E_{2^+}$ energies 
and (b) summed  $B(E2)$  strength (filled squares) and $B(E2, 0^+_{\rm g.s.}
 \rightarrow 2^+_1)$ 
values 
with the experimental data. The summed $B(E2)$ strength  includes all the 
transitions up
to an excitation energy of 9 MeV.
}
 \label{fig:qrpa_be2}
\end{center}
\end{figure}

While the SMMC approach allows for the calculation of the summed 
strength in large model spaces, it is not capable of making detailed 
spectroscopic predictions. For this reason, we have also performed QRPA 
and  diagonalization shell-model calculations.
Our QRPA calculations closely follow the formalism described recently in
 Ref.~\cite{Terasaki02}.
As a residual two-body interaction, we use the sum of an isoscalar 
and an isovector quadrupole force, and a quadrupole
pairing force.
For the single-particle levels  below the 
$N$ $(Z)$ = 82 shell gap. we took those of the Woods-Saxon potential
\cite{[Cwi87]},
 and  the unbound states were
approximated by  the Nilsson levels \cite{[Nil69]}. 
Guided by the experimental data (cf. Ref.
\cite{[Boh69]}), the energy of the $2p_{3/2}$ neutron state was
shifted up by 1 MeV.
For the  strength of the isoscalar quadrupole force, we adopted
the self-consistent value multiplied by 0.8, and for the isovector force
we took 
$\chi_{T=1} = -\frac{123.8}{A^{7/3}}$ [$\frac{\rm MeV}{\rm fm^4}$]. 
The renormalization factors of the pairing gaps are 0.8 (neutron) and 0.9 
(proton). 
The bare charges are used for calculation of 
$B(E2;0^+\rightarrow 2^+)$, consistent with the large model space. 
The results are shown in Fig. \ref{fig:qrpa_be2}. 

\begin{figure}[ht]
\begin{center}
    \leavevmode
    \includegraphics[width=0.80\columnwidth]{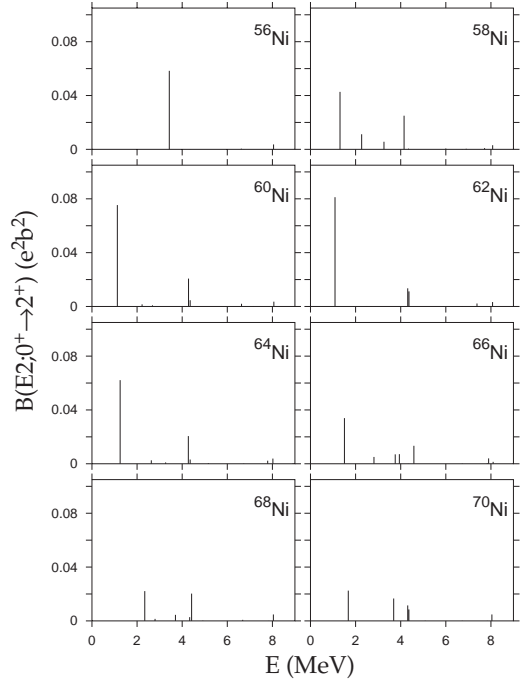}
\caption{Distribution of the
$B(E2; 0^+_{\rm g.s.} \rightarrow 2^+)$ strength for  even-even  Ni isotopes
calculated in the QRPA method.}
\label{fig:qrpa_str}
\end{center}
\end{figure}

Our calculations nicely reproduce the observed
trend of the $E_{2^+_1}$ energies in Ni
isotopes,  including the pronounced rise at $^{56}$Ni and $^{68}$Ni.
The QRPA calculations also give a reasonable description of  the 
$B(E2, 0^+_{\rm g.s.} \rightarrow 2^+_1)$ values, 
with the maximum  around $^{62}$Ni and the strong decrease towards $^{68}$Ni.
The structure of the  lowest 2$^+$ QRPA phonon  in 
$^{68}$Ni is dominated by neutrons (90\%).
Importantly,
 our QRPA calculations confirm that most of the $B(E2)$
strength in $^{68}$Ni resides in excited states, in contrast to
$^{56}$Ni, where the $B(E2)$ strength is exhausted by the transition to
the first $2^+$ state. These arguments are demonstrated again in 
Fig.~\ref{fig:qrpa_be2}, which displays the summed $B(E2)$ strength (filled squares)
and in
Fig.~\ref{fig:qrpa_str},
which shows the predicted  $B(E2)$ strength function. It is seen that
around neutron number $N=40$, a  significant quadrupole strength develops
at excitation energies around 4 MeV.

Finally, our arguments have been tested and confirmed in
large-scale diagonalization shell-model 
calculations of the $B(E2)$ strength distribution in $^{66,68}$Ni.
We adopt the same valence space and the
effective interaction {\it fpg} employed in Ref.~\cite{Sorlin02}.
This valence space consists of a $^{48}$Ca core
(more precisely, a $^{40}$Ca core with eight $f_{7/2}$ frozen
neutrons), the $f_{7/2}$, $p_{3/2}$, $p_{1/2}$, and $f_{5/2}$ active
orbitals for protons and the $p_{3/2}$, $p_{1/2}$, $f_{5/2}$, and
$g_{9/2}$ active orbitals for neutrons.   
The SM  calculations 
describe well the behavior of the $2^+$ energies and
$B(E2)$ rates  for the Ni isotopes ranging from N=28 to N=40 \cite{Sorlin02}. 
\begin{figure}[ht]
\begin{center}
    \leavevmode
    \includegraphics[width=0.70\columnwidth]{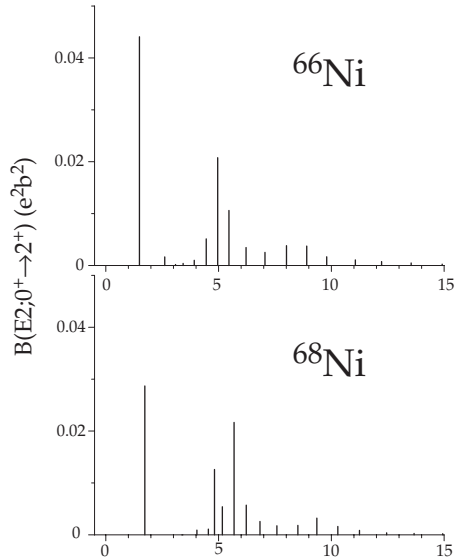}
\caption{Distribution of the
$B(E2; 0^+_{\rm g.s.} \rightarrow 2^+)$  strength in  $^{66,68}$Ni 
calculated in the 
diagonalization shell model.}
\label{fig:shell}
\end{center}
\end{figure}
The $B(E2)$ strength distribution shown in Fig.~\ref{fig:shell} has been calculated 
at a truncation
level which considered up to a total of 7 particle excitations from the
$f_{7/2}$ orbital to the upper $fp$ shell for protons and from the upper
$fp$ shell to the $g_{9/2}$ orbital for neutrons.
 For $^{68}$Ni, the calculation predicts a
$B(E2)$ value of about 280 e$^2$fm$^4$, which nicely
agrees  with the experimental value. However, the transition
to the first $2^+$ state exhausts only the smaller fraction of the total
shell-model $B(E2)$ strength, which we calculate as 900 e$^2$fm$^4$.
Most of the calculated strength resides at excitation
energies around 5-6 MeV. 
The  wave function of the first $2^+_1$ state involves 
mainly (2p-2h) neutron excitations; 
the proton configuration is a mixture of 0p-0h
($50\%$), 1p-1h (($25\%$),  and 2p-2h ($20\%)$ components. The SM
$^{68}$Ni ground state  corresponds to a closed-shell configuation
plus a $35\%$ admixture of 2p-2h neutron excitations. The results 
for $^{66}$Ni nicely confirm the QRPA prediction: the transition strength is gradually
shifted from the first excited state to higher energies as approaching  $N$=40.

In summary, we have performed microscopic calculations
of the $B(E2;0^+_{\rm g.s.} \rightarrow 2^+)$ strength distribution
in $^{68}$Ni, and in other even-even nickel
isotopes. Our main finding is that a significant portion of the $B(E2)$
strength in $^{68}$Ni resides in excited states around 5 MeV, and that
the small observed $B(E2, 0^+_{\rm g.s.} \rightarrow 2^+_1)$
 value is not necessarily
an argument for a shell closure at $N=40$.
 In fact, we argue that this
transition rate  is quite sensitive to the energy splitting between $fp$ shell and
$g_{9/2}$ orbital, and that its smallness might indeed be an indication
for a rather small gap.  In short, the small $B(E2)$ 
rate reflects the fact that the lowest $2^+$ state in
$^{68}$Ni is primarily a neutron excitation (cf. Ref.~\cite{Terasaki02}
for a similar discussion for $^{136}$Te).
If our arguments are correct, then most of the $B(E2)$ strength should
 reside in excited states. Further experimental investigations, including the
 $g$-factor measurement in $^{68}$Ni, are certainly called for.

\begin{acknowledgments}
We are indebted to Jonathan Engel  for the use of his QRPA code.
This work was supported in part by the U.S.~Department of Energy under Contract
Nos.~DE-FG02-96ER40963 (University of Tennessee), DE-AC05-00OR22725 with UT-Battelle,
LLC (Oak Ridge National Laboratory), 
 the National Science Foundation Contract
 No.~0124053 (U.S.-Japan Cooperative Science Award), and by the Danish Research
 Council. Computational resources were
 provided by the Center for Computational
 Sciences, ORNL, and the National Energy Research Scientific Computing Center, Berkeley.
\end{acknowledgments}


\end{document}